\documentclass[conference]{IEEEtran}
\IEEEoverridecommandlockouts
\usepackage{cite}
\usepackage{amsmath,amssymb,amsfonts}
\usepackage{algorithmic}
\usepackage{graphicx}
\usepackage{textcomp}
\usepackage{xcolor}
\usepackage{amssymb}
\usepackage{booktabs} 

\usepackage{algorithm}
\usepackage{algorithmic}

\usepackage[colorlinks=true,linkcolor=black,urlcolor=magenta,citecolor=black]{hyperref}

\def\BibTeX{{\rm B\kern-.05em{\sc i\kern-.025em b}\kern-.08em
    T\kern-.1667em\lower.7ex\hbox{E}\kern-.125emX}}
\begin{document}

\title{
A$^3$PIM: An Automated, Analytic and Accurate Processing-in-Memory Offloader
}


\newcommand{\summary}[1]{\textcolor{blue}{#1}}
\newcommand{\confused}[1]{\textcolor{red}{#1}}
\newcommand{\diffprof}[1]{\textcolor{black}{#1}}
\newcommand{\todo}[1]{\textcolor{magenta}{#1}}
\newcommand{\jiang}[1]{\textcolor{purple}{#1}}

\author{\IEEEauthorblockN{Qingcai Jiang$^*$, Shaojie Tan$^*$, Junshi Chen and Hong An}
\IEEEauthorblockA{School of Computer Science and Technology, University of Science and Technology of China, Hefei, China \\
}
\IEEEauthorblockA{Email: \{jqc, shaojiemike\}@mail.ustc.edu.cn, 
\{cjuns, han\}@ustc.edu.cn}
}

\maketitle
\def\thefootnote{*}\footnotetext{These authors contributed equally to this work.}\def\thefootnote{\arabic{footnote}}
\begin{abstract}

The performance gap between memory and processor has grown rapidly.
Consequently, the energy and wall-clock time costs associated with moving data between the CPU and main memory predominate the overall computational cost.
The Processing-in-Memory (PIM) paradigm emerges as a promising architecture that mitigates the need for extensive data movements by strategically positioning computing units proximate to the memory.
Despite the abundant efforts devoted to building a robust and highly-available PIM system, identifying PIM-friendly segments of applications poses significant challenges due to  the lack of a comprehensive tool to evaluate the intrinsic memory access pattern of the segment. 

To tackle this challenge, we propose A$^3$PIM\footnote{The code and benchmarks of this work are opened-sourced in: https://github.com/ACSA-PIM/A3PIM}: an Automated, Analytic and Accurate Processing-in-Memory offloader. We systematically consider the cross-segment data movement and the intrinsic memory access pattern of each code segment via static code analyzer. 
We evaluate A$^3$PIM across a wide range of real-world workloads including GAP and PrIM benchmarks and achieve an average speedup of 2.63x and 4.45x (up to 7.14x and 10.64x) when compared to CPU-only and PIM-only executions, respectively. 

\end{abstract}

\begin{IEEEkeywords}
Processing-in-Memory,
Static Analysis,
Workload Offloading
\end{IEEEkeywords}

\section{Introduction}
The widening discrepancy between the speeds of the CPU and main memory system has emerged as a major issue in modern computer systems, colloquially termed as the "memory wall". In demonstrating this, the time required for one off-chip memory access from the CPU is approximately 100 times that of a double-precision operation, with energy consumption being about 1000 times greater. Simultaneously, recent years have seen the rise of data-intensive applications such as graph processing and machine learning inference, which expend a significant amount of time and energy transferring data between the CPU and main memory. As a result, this phenomenon of data movement has become a principal concern in contemporary computing systems. 

\begin{figure}[htbp]
    \centering
    \centerline{\includegraphics[width=\linewidth]{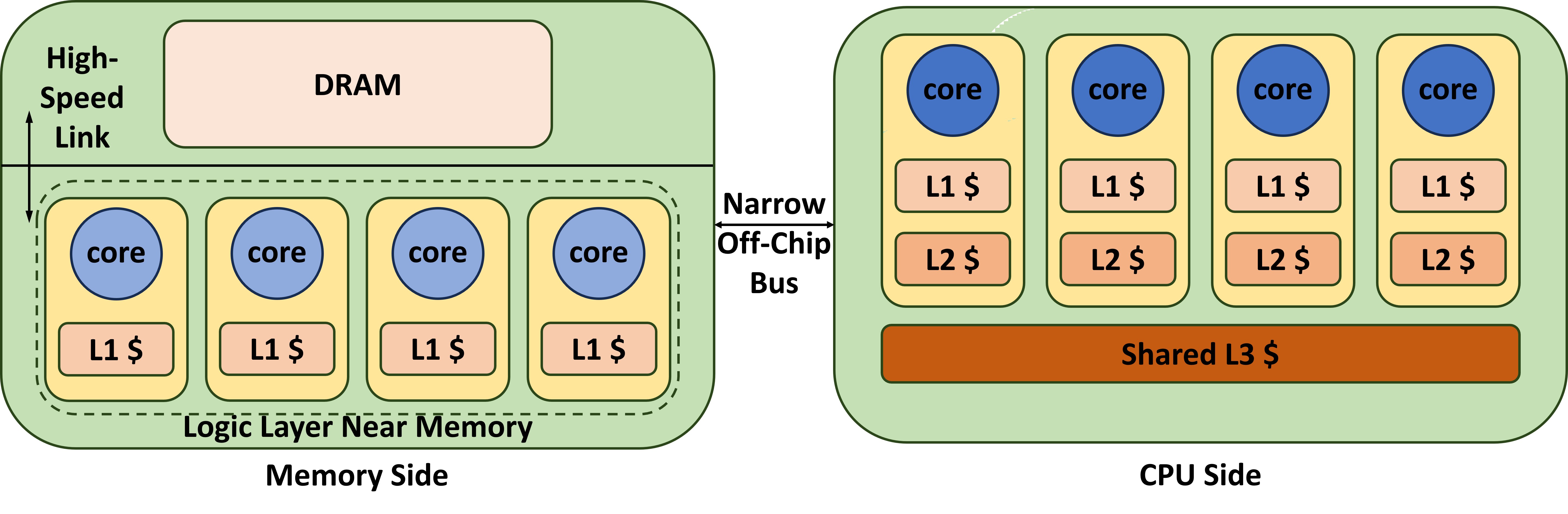}}
    \caption{High-level PIM architecture}
    \label{fig:pim-arch}
\end{figure}

Fortunately, the advent of 3D-stacked memory technologies has heralded the dawn of Processing-in-Memory (PIM) as a promising solution to counter the data movement problem. 
These PIM architectures integrate computational units within the logic layer of 3D-stacked memory, strategically positioning these units close to the data to reduce data movements.
Fig.~\ref{fig:pim-arch} depicts an typical PIM architecture, where computing units are integrated close the memory side, such as the logic die of Micron's hybrid memory cube (HMC)~\cite{hmc}. The computing units are able to perform logic operations with ultra-high memory access bandwidth. In such a CPU-PIM system, there are two different types of processing units that are suitable for different kinds of workloads: PIM units specialize in workloads with intensive and irregular memory access patterns, while CPUs excel at handling workloads with regular memory access patterns that can be effectively cached.

To extract the benefits of such CPU-PIM system, an intelligent offloader must be developed to choose the best-fitting core for each segment of a workload. 
However, exploiting such an offloader remains a challenge as multiple factors must be jointly taken into account: 
1. The compatibility of the segment to the PIM and CPU core in terms of the memory access pattern, as previously discussed. 
2. The inter-segment switching cost which involves data movement between PIM and CPU because of data dependency, and the context-switching cost between PIM and CPU threads.
ALP~\cite{ghiasi2022alp} tackle this challenge by introducing a novel prefetcher to alleviate data movement overhead. PIMProf~\cite{wei2022pimprof} answer this question
by implementating an automated PIM profiling and offloading tool. Additionally, many previous works aim to offload predetermined segments that have potential to be accelerated by PIM, such as special instructions~\cite{hsieh2016transparent,ahn2015pim} and function kernels~\cite{ahn2015scalable}. Despite this, a significant shortcoming of these research efforts is the absence of a holistic tool to assess the intrinsic memory access pattern of each code segment, which is a critical determinant in deciding the offloading policy of the CPU-PIM system.

In this work, we design A$^3$PIM, an automated, analytic and accurate PIM offloader for CPU-PIM system.
The contributions of this work are the following:
\begin{itemize}
    \item We identify and characterize the three key factors in the offloading of PIM segments: 1. The compatibility of the segments to PIM and CPU core, 2. The inter-segment data movement overhead between the PIM and CPU cores when the application is partitioned between them.and 3. The context-switching time between PIM and CPU threads.
    \item We design a novel basic clock offloader on CPU-PIM system, A$^3$PIM, which leverage static code analyzer to evaluate the intrinsic memory access pattern of each basic block and design an novel algorithm that jointly consider the compatibility of the segment to the PIM and CPU core and data movement overhead.
    \item We evaluate the effictiveness of A$^3$PIM on a wide range of workloads: GAP~\cite{beamer2015gap} and PrIM~\cite{gomez2021benchmarking}. The nummerical results show that the CPU-PIM hybrid execution by A$^3$PIM achieve an average speedup of 2.63x and 4.45x (up to 7.14x and 10.64x) when compared to CPU-only and PIM-only executions, respectively.
\end{itemize}



\section{Background And Motivation}

\subsection{Related Work}


Prior workds on PIM offloading exhibits two significant limitations. 
Firstly, some existing works necessitate additional hardware implementations for runtime scheduling~\cite{ghiasi2022alp,wei2022pimprof}, resulting in time-consuming investigations and added hardware design complexity. 
Secondly, previous studies predominantly concentrate on specific segment properties, such as data locality~\cite{ahn2015pim} or cache hit rates~\cite{gao2015practical} during runtime measurements. However, these approaches lack static intrinsic metrics for scheduling units, which could unveil segment affinity for allocation to either the CPU or PIM.


\subsection{Static Code Analyzer}

A static code analyzer~\cite{iaca2017,lattner2004llvm} is a tool designed to investigate the intrinsic properties of a given \textbf{basic block}, such as the estimated execution time, memory access patterns, and dependencies between each instruction.
Previous works~\cite{jiang2022quantifying,tan2023uncovering,mendis2019ithemal} about static code analyzers have unveiled the intrinsic properties of code basic blocks, including memory port pressure. This analysis has shed light on the performance preferences of code basic blocks when executed on either PIM or CPU cores, which can serve as a theoretical foundation for decision-making.


\subsection{Data Movement Overhead}

During the offload process, offloading an application's code into segments on PIM or the CPU can lead to a substantial amount of inter-segment data movement overhead, i.e., data generated in one segment and used in other segments.
Previous research~\cite{suleman2010data} has demonstrated that the instructions responsible for generating inter-segment data remain consistent across various executions of a program with different input sets. This consistency ensures that we can make offloading decisions statically, without the need for dynamic offloading.



\section{Cost Modeling of A$^3$PIM}


\subsection{Componets of execution time in CPU-PIM system}
In order to provide good offloading decisions for a program, it is necessary to understand the cost model of different program executions for minimizing the total cost. 
We identify \textbf{two major sources of cost}: 
1. The execution cost which is the execution time of the code region on CPU or PIM and 
2. The switching cost which is the overhead for maintaining the consistency of data and the program context when switching between CPU and PIM.

1) Execution cost: 
In the offload decision process, a key concern is that each segment of a program varies in execution time when running on either a CPU or a PIM core.
This numerical difference arises from architectural disparities between the PIM and CPU core, such as memory access latency, which can be statically evaluated using the inherent characteristics obtained from a static code analyzer.

2) Switching cost: The switching cost comes from two sources: data dependency and context switch.

The \textbf{first} source comes from the data dependency between code regions that are placed on different processing units, e.g., one region on CPU and another on PIM. In this scheme, the data must be transferred from one core's cache to another's. We succinctly refer to this process as cache line data movement (CL-DM).

\begin{figure}[htbp]
    \centering
    \centerline{\includegraphics[width=0.85\linewidth]{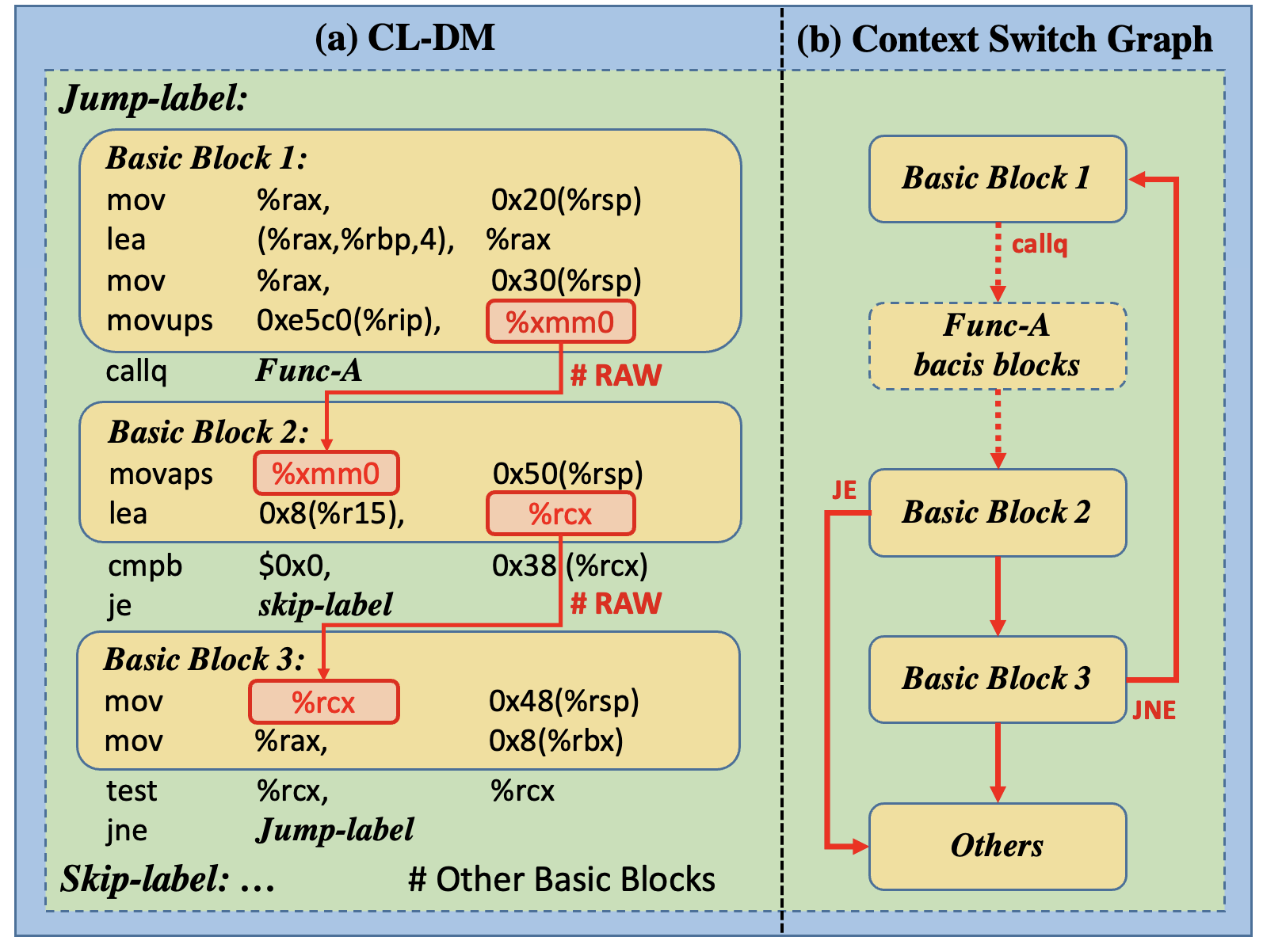}}
    \caption{Example of (a) Cache Line Data Movement (CL-DM) and Corresponding (b) Context Switch Graph in Continuous ARM64 Assembly Code.}
    \label{fig:cl}
\end{figure}

\textbf{CL-DM.} We analyze the data dependency between basic blocks at cache line granularity, as memory transfer is cache-line-grained. When the same cache line of data is shared by two program regions that execute in different places (PIM and CPU), we model a single data transfer as the total cost of one cache line flush issued by the source, plus one cache line fetch issued by the destination. Figure~\ref{fig:cl} shows a code example with multiple basic blocks. Assume that we execute $basic\ block\ 1$ , $basic\ block\ 3$ on CPU and $basic\ block\ 2$ on PIM, 
then when we switch between them, the memory locations of  $\%xmm0$ and $\%rcx$ each incurs a data dependency cost.
The number of data dependency instances increases if there are multiple shared locations. To compute the total data dependency cost of a certain offloading decision, a naive way is to go over all memory accesses, and increase the cost wherever a data transfer happens. 

The \textbf{second} source comes from the context switch, which mainly includes the overhead of saving and restoring the process or thread states, TLB information etc. Different from the data dependency cost, context movement has a more or less constant cost, which is determined by the operating system.
\diffprof{We refer to this cost as light-weight context switch data movement.}
Previous research has either overestimated the cost of a context switch to 2000 cycles~\cite{wei2022pimprof} or underestimated it to 0 cycles~\cite{ghiasi2022alp}.
In our work, we estimate the context switch cost for CPU-PIM to be around 800 cycles.

This estimated cost is derived from the following factors:

\begin{enumerate}
    \item Given the current programming models, the context switch between PIM and CPU takes place at the thread level, not the process level. As a result, the context switch between threads on PIM and CPU is anticipated to have a more streamlined context, thereby reducing the overhead. 
    \item PIM cores generally make use of Arm cores with L1 caches that adhere to the Virtually Indexed, Physically Tagged (VIPT) scheme~\cite{ARM-VIPT}. During thread switching, there is no need to alter the virtual address mapping, thus eliminating the need to flush the L1 cache. Consequently, this further lowers the overhead associated with context switching.
\end{enumerate}

To further validate our estimation, we conduct a thread context switching cost test on a real 64-core Armv8 server processor~\cite{kunpeng920} following the experimental steps outlined in~\cite{litton2016light}. The result approximates to around 800 cycles. Thus, based on these considerations, we adopt an assumed context switch cost of 800 cycles for the CPU-PIM system.

\textbf{Context Switch Cost.} The context switch cost arises when two neighboring basic blocks are executed in different cores.
Typically, the context switch cost is constant and depends on the operating system. To compute this cost, A$^3$PIM keeps track of the program's execution as it crosses the boundary between basic blocks. This is achieved by constructing a weighted directed graph, where each edge's weight represents the number of times the execution transitions from one region to another. Figure~\ref{fig:cl} shows an example of the context switch graph.  




To elaborate on the cost modeling in the CPU-PIM system, we design an experiment to analyze the proportion of each time cost component across a range of applications, as illustrated in Table~\ref{tbl:clVSreg}. The corresponding simulation methodologies will be detailed in~\ref{sec:evaluation}. 
We make a key observation that when scheduling is performed at the granularity of basic blocks, the cost of context switch is significantly larger than the cost of CL-DM. 
This observation provides valuable guidance for our design of data movement overhead elimination, emphasizing the importance of addressing context switch to achieve efficient execution in the system.

\begin{table}[htbp]
    \centering
    \caption{Comparison of different applications' execution time and data movement time using Architecture-Suitability/Greedy offload strategy}
    \begin{tabular}{lrrr}
        \toprule
         &  &  \multicolumn{2}{c}{\textbf{Data Movement Time}} \\ 
        \textbf{Application} & \textbf{Execution Time} & \textbf{CL-DM} & \textbf{Context Switch}\\ 
        \midrule
        bc &31.37\%                     &14.17\%                     &54.46\%                         \\
sssp &1.56\%                     &1.57\%                     &96.86\%                         \\
bfs &49.59\%                     &2.21\%                     &48.2\%                         \\
pr &71.74\%                     &0.01\%                     &28.24\%                         \\
select &8.82\%                     &0.0\%                     &91.18\%                         \\
unique &10.62\%                     &0.0\%                     &89.37\%                         \\
        \textbf{AVERAGE} &\textbf{28.95\%}& \textbf{3.0\%}& \textbf{68.05\%}\\
        \bottomrule
    \end{tabular}
    \label{tbl:clVSreg}
\end{table}

While measuring the cost of CL-DM necessitates dynamic tracking of memory traces and cache line usage\cite{ghiasi2022alp,wei2022pimprof}, our observation regarding the relationship between context switch cost and CL-DM suggests that we can achieve CL-DM elimination as a byproduct of context switch cost elimination. This revelation allows us to develop a comprehensive set of static methods for eliminating both context switch cost and CL-DM which will be detailed introduced in Section~\ref{DM-Removed}.

\subsection{Cost modeling formula}


Based on the above observations and discussions, we propose the following equation to model the time overhead of scheduling in a CPU-PIM system.

\begin{footnotesize}
\begin{align*}
Time\ Overhead &= \sum_{i \in PIM}PIM_i +  \sum_{j \in CPU}CPU_j \\&+\sum_{i \in PIM}\sum_{j \in CPU}(CL\_DM(i,j)+CXT(i,j))
\end{align*}
\end{footnotesize}

Here, $CL\_DM(i,j)$ represents the cost of cache line data movement generated when program region i is scheduled on PIM and program region j is scheduled on the CPU. 
Similarly,  $CXT(i.j)$ represents the cost of context switch data movement incurred when unit i is scheduled on PIM and unit j is scheduled on the CPU. 
Additionally, $PIM_i$  denotes the execution time of program region i when scheduled on PIM, and  $CPU_j$ denotes the execution time of program region j when scheduled on the CPU. 
Both $PIM_i$ and $CPU_j$ can be collectively referred to as the execution time. 
Furthermore, $CL\_DM(i,j)$ and $CXT(i,j)$ can be commonly referred to as data movement overhead.
This formulation effectively captures the intricate interplay of data movement costs and execution times in the CPU-PIM system, providing a comprehensive model for analyzing scheduling overhead.



\begin{figure*}[htbp]
    \centering
    \centerline{\includegraphics[width=.9\linewidth]{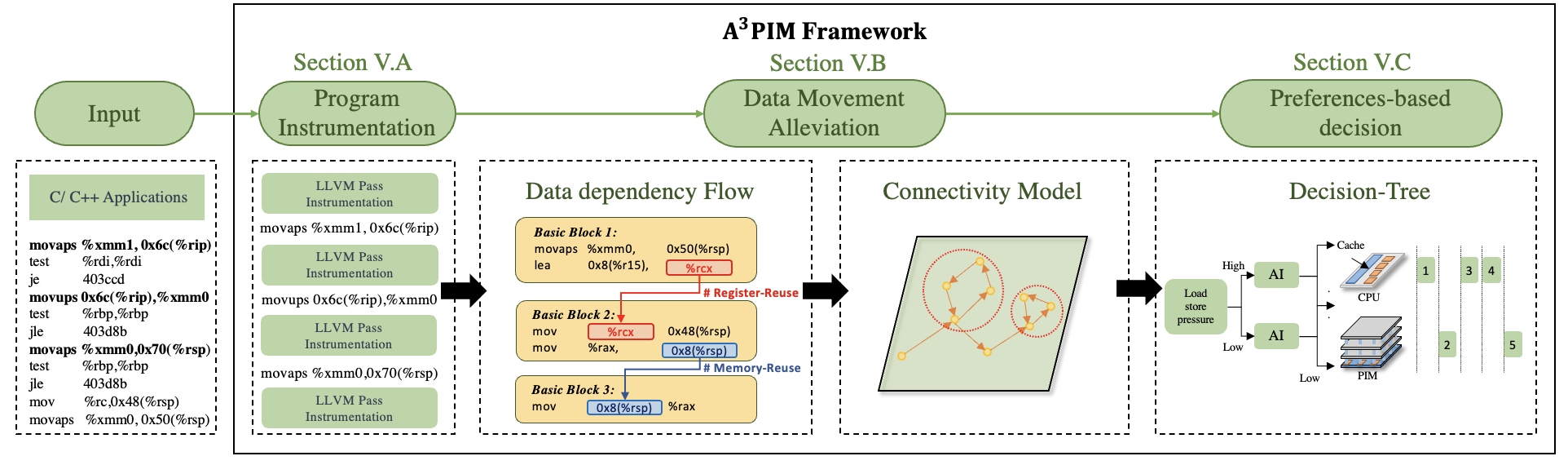}}
    \caption{A$^3$PIM overview}
    \label{fig:workflow}
\end{figure*}

In previous works, there have been two main approaches to minimize the overall cost in the above formula. 
Some works~\cite{hsieh2016transparent,nai2017graphpim} focused on measuring partial values, such as optimizing for overall memory bandwidth savings, where the actual data movement overhead was measured for scheduling purposes. 
However, these approaches overlooked the benefits in terms of execution time. 
Other works~\cite{wei2022pimprof} considered both the execution time of scheduling units and the data movement overhead between them. 
They designed heuristic algorithms to reduce the time complexity of traversing all possible scheduling combinations, as exhaustively exploring all combinations would have a time complexity of $2^N$ for N scheduling units.

We make a key observation that the data movement overhead is significantly larger than the execution time. Inspired by this, we propose a two-stage minimization algorithm. The first stage involves clustering to eliminate data movement overhead, and the second stage leverages the intrinsic characteristics of scheduling units to find the scheduling choices with the smallest execution time. By doing so, we obtain near-optimal scheduling results without relying on dynamic measurement information. 

This approach effectively addresses the challenge of minimizing the overall cost in the system without dynamically measuring information, taking into account the significant impact of data movement overhead compared to execution time.

\section{A$^3$PIM Design}
\label{A3PIM}
Based on above observations, we present A$^3$PIM, an novel end-to-end compile-time scheduling framework designed for efficient task distribution between the CPU and PIM architectures. The overall workflow of A$^3$PIM is shown in Figure~\ref{fig:workflow}.


\subsection{Program Instrumentation}

A$^3$PIM statically instruments the program using an LLVM~\cite{lattner2004llvm} compiler pass. 
It divides the entire program into small regions to enable a fine-grained profiling by inserting lightweight marker functions, and the program regions are used as the basic unit for profiling and offloading. 
The granularity of program regions is configurable so that applications that vary in size, parallelism and data dependency patterns may all benefit from offloading.
In this work, we compare the different granularities of regions within a program, specifically basic blocks~\cite{allen1970control} and functions.


\subsection{Data Movement Alleviation}
\label{DM-Removed}

In the first step of A$^3$PIM algorithm, based on automatically-instrumented region boundaries (e.g., basic block or function), we aim to cluster related regions in order to reduce the impact of high inter-segment data movement. 
Taking into account the two types of data movement, we employ a comprehensive clustering method that considers various factors to mitigate the cost of data movement.

Based on previous research, we integrate two methods: one based on memory access temporal locality metrics\cite{locality2005sc} and the other on register data dependency between basic blocks\cite{ghiasi2022alp}, aiming to eliminate data movement. Combining these methods, we propose the following formula for connectivity:

\begin{footnotesize}
\[ 
\text{Connectivity} = \frac{\alpha \cdot \text{Memory\_Reuse} + (1 - \alpha) \cdot \text{Register\_Reuse}}{\text{Instruction\_Count}}
\]
\end{footnotesize}


The connectivity metric calculates the tightness of data movement between two basic blocks, serving as the basis for clustering.

Here, $\alpha$ is a parameter that balances the contributions of memory access temporal locality and register data dependency to data movement overhead. 
$Memory\_Reuse$ represents the number of times two basic blocks access the same memory address, while $Register\_Reuse$ counts the occurrences of accessing the same register between two basic blocks. 
$Instruction\_Count$ denotes the maximum number of instructions among the two basic blocks. 
Introducing this parameter is essential because, in situations with identical $Memory\_Reuse$ and $Register\_Reuse$, a higher number of instructions is more likely to hide data movement latency, and greater instruction count may exceed the cost of scheduling to different locations.

This formula allows us to flexibly adjust the relative importance of both methods, enabling us to achieve optimal data movement elimination results under various scenarios.

A connectivity value of 0 indicates no data reuse, while a value close to 1 suggests a very high data movement cost (i.e., a value near 1 implies that the instructions in basic blocks almost exclusively contain reused memory addresses or registers).

After clustering based on the connectivity value, all the regions of the program are grouped together with their tightly-connected neighboring regions.
This clustering effectively reduces the data movement between these blocks, leading to significant alleviation of data movement overhead.

\subsection{Preferences based on intrinsic characteristics of clusters}
\label{exeCost}

After the clustering step, A$^3$PIM collects static statistics for each cluster using a static code analyzer. These statistics include the predicted number of execution cycles, the load-store port pressure for each cluster, and the quantity and type of instructions in each cluster.
We utilize the flollowing static parameters to obtain the optimal offload decision for each cluster:


\begin{itemize}
    \item \textbf{Load-store port pressure} is a outcome of a static code analyzer, indicating the level of congestion on the load-store ports during the execution of instructions within a basic block. 
    This information provides crucial insights into the characteristics and resource requirements of each region.
    \item \textbf{Arithmetic intensity (AI)} is defined as the ratio of computational instructions to memory access instructions within a scheduling unit. This metric quantifies the amount of computation performed per memory request.
\end{itemize}

Different from previous works, we do not utilize the last-level CPU cache misses per Kilo-instruction (MPKI) metric since we consider it would introduce hardware overhead to performance monitor counter and runtime scheduling overhead. Load-store port pressure is capable of fully replacing its functionality.

Algorithm~\ref{alg1} represents the key takeaways we obtain from our memory bottleneck classification. Based on three key metrics, we classify workloads into those that prefer execution on PIM and those that prefer execution on the CPU.


\begin{algorithm}
 \renewcommand{\algorithmicrequire}{\textbf{Input:}}
 \renewcommand{\algorithmicensure}{\textbf{Output:}}
 \caption{Offloading Clusters With the instrinsic characteristics of clusters}
 \label{alg1}
 \begin{algorithmic}[1]
     \REQUIRE the basic blocks of the cluster
     \ENSURE The mapping of each cluster to PIM or CPU

    \IF{The cluster shows high parallelism}
        \STATE MAP(cluster, PIM);
    \ELSE
        \IF{The cluster suffer in load-store port pressure}
            \STATE MAP(cluster, PIM);
        \ELSE
            \IF{The cluster shows high memory intensity}
                \STATE MAP(cluster, PIM);
            \ELSE
                \STATE MAP(cluster, CPU);
            \ENDIF
        \ENDIF
    \ENDIF
 \end{algorithmic}  
 
\end{algorithm}



In summary, our work shows that employing a static code analyzer can provide sufficient support for PIM scheduling based on static information. 
By combining cost clustering of data movement in advance, we can effectively eliminate a significant portion of data movement overhead.

\section{Evaluation Methodology}
\label{sec:evaluation}


\subsection{Evaluation Configurations}

We model a CPU-PIM system on the Sniper simulator~\cite{sniper_carlson2014evaluation}. The configuration of this system is listed in Table~\ref{tab:sys_conf}. The baseline configuration comprises out-of-order CPU cores, while the PIM configuration includes atom-like in-order cores~\cite{ahn2015scalable}.

\begin{table}[htbp]
    \centering
    \caption{SYSTEM CONFIGURATION}
    \begin{tabular}{|c|}
        \hline
        \textbf{Out-of-Order CPU (baseline)}\\
        \hline
        1 General purpose processors \\
        3GHz, 4-way superscalar \\
        32kB L1I, 32kB L1D, 256kB L2, 2MB L3\\
        \hline
        \textbf{General-purpose in-order (PIM)}\\
        \hline
        32 general purpose cores \\
        32kB L1I, 32kB L1D \\
        \hline
        \textbf{Data movement Cost} \\
        \hline
        Cache line fetch/flush on CPU: 60 ns, on PIM: 30 ns \\
        Register data movement: 2 cache line fetch \& flush \\
        \hline
    \end{tabular}
    \label{tab:sys_conf}
\end{table}

\begin{figure*}[thbp]
    \centering
    \includegraphics[width=0.9\linewidth]{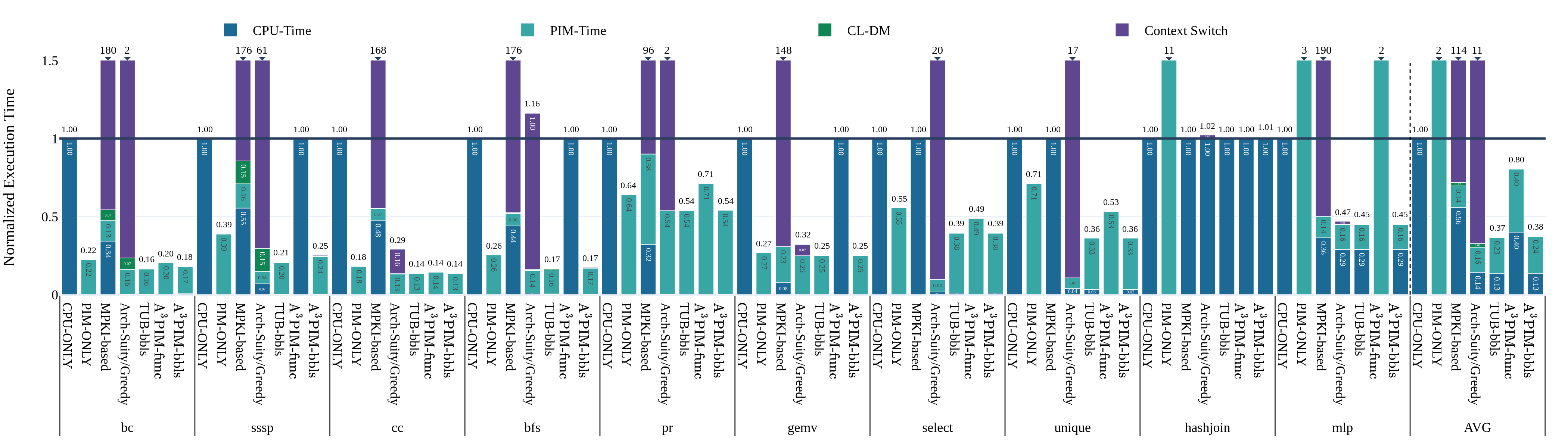}
    \caption{Execution time breakdown of GAP and PrIM workloads using different offloading decisions}
    \label{fig:result}
\end{figure*}

\subsection{Evaluation Workloads}

For evaluation, We use two widelyused benchmark suites that contain a variety of workloads to demonstrate the flexibility of A$^3$PIM: (1) graph benchmark suite (GAP)~\cite{beamer2015gap} — high memory intensity and parallelism (bc, sssp, cc, bfs and pr). (2) PrIM benchmarks~\cite{gomez2021benchmarking} - the first benchmark suite for a real-world PIM architecture (gemv, select, unique, hashjoin and mlp).

\section{Evaluation}

\subsection{System Configurations} 

In our evaluation, we consider five distinct system configurations as the baseline models for comparison against A$^3$PIM.

\begin{itemize}
    \item \textbf{CPU-only} refers to offload all regions to CPU.
    \item \textbf{PIM-only} refers to offload all regions to PIM.
    \item \textbf{MPKI-based} refers to offload the regions to PIM if the MPKI of the program region exceeds the threshold.
    \item \textbf{Architecture-Suitability/Greedy}  refers to offload to the min execution-cost location without considering the data dependency cost.
    \item \textbf{Theoretical Upper Bound (TUB)} refers to the minimum value obtained by exploring all possible permutations of scheduling combinations.
    \item \textbf{A$^3$PIM: Data-movement-aware \& intrinsic-characteristics-aware}: 
    A$^3$PIM will undergo evaluation with two distinct offloading granularities: \emph{function-level scheduling granularity} (A$^3$PIM-func) and \emph{basic-block-level scheduling granularity} (A$^3$PIM-bbls).
\end{itemize}

\subsection{Performance}
\label{result}

Figure~\ref{fig:result} depicts the execution time breakdown for each application using six different offloading strategies.

We make four key oberservations from these results:
\textbf{First}, not all applications benefit from complete offloading to PIM. Graph applications and memory-intensive applications such as gemv, select, and unique demonstrate significant improvements, while hashjoin and mlp underperform, resulting in a speed decrease in PIM-ONLY strategy on average.
\textbf{Second}, both the traditional MPKI and Greedy methods suffer from substantial data movement overhead, primarily attributed to Context Switch costs rather than CL-DM. However, the Greedy method proves suitable for specific applications like cc, gemv, and mlp, which consist of multiple independent kernels.
\textbf{Third}, A$^3$PIM-bbls is performing better than A$^3$PIM-func. A$^3$PIM-func provides a 1.25× speedup over CPU-only and a 2.11× speedup over PIM-only, while A$^3$PIM-bbls offers a 2.63× speedup over CPU-only and a 4.45× speedup over PIM-only. This performance difference arises from the coarse partitioning introduced by function-level granularity, leading to a loss in the potential performance gains achievable through finer-grained scheduling.
\textbf{And last}, A$^3$PIM-bbls effectively offloads all application cases, including PIM-friendly graph applications, CPU-friendly applications like hashjoin, and even uncovers the PIM benefit potential in specified applications such as mlp, where the PIM-ONLY method leads to performance degradation. Consequently, A$^3$PIM-bbls approaches the theoretical peak attainable performance speedup of 4.56x over PIM-ONLY.



\section{Conclusion}

 In this work, we proposes A$^3$PIM, an automated, analytic and accurate offloading framework for CPU-PIM systems. By leveraging static code analysis to model the intrinsic characteristics of code segments, A$^3$PIM is able to make informed task partitioning and scheduling decisions to minimize data movement overheads. Our evaluations on real-world benchmarks demonstrate that A$^3$PIM can effectively extract the performance benefits of the PIM architecture and achieve average of 2.63x and 4.45x speedups compared to CPU-only and PIM-only executions. The proposed techniques provide a promising direction to enable performance-aware scheduling for heterogeneous CPU-PIM systems without costly runtime profiling. 

\section*{Acknowledgment}
This work is partly supported by Strategic Priority Research Program of the Chinese Academy of Sciences (Grant No. XDB0500102) and the National Natural Science Foundation of China (Grant No. 62102389).








\end{document}